\documentclass[review]{elsarticle}

\usepackage{hyperref}

\journal{Journal of \LaTeX\ Templates}

\begin{document}

\begin{frontmatter}

\title{On concentration of $^{42}$Ar in the Earth's atmosphere}



\author[ITEP]{A.S. Barabash\corref{mycorrespondingauthor}} 



\ead{barabash@itep.ru}


\author[UCL]{R.R. Saakyan}


\author[ITEP]{V.I. Umatov}


\cortext[mycorrespondingauthor]{Corresponding author}


\address[ITEP]{National Research Center ``Kurchatov Institute'', Institute of Theoretical

 and Experimental Physics, B. Cheremushkinskaya 25, 117218 Moscow, Russia}

\address[UCL]{University College London, London WC1E 6BT, United Kingdom}

\begin{abstract}


Data from the DBA  liquid argon ionization chamber experiment have been used
to obtain an estimate on the concentration of $^{42}$Ar in the Earth's
atmosphere, $6.8^{+1.7}_{-3.2}\cdot10^{-21}$ atoms of $^{42}$Ar per atom of $^{40}$Ar
corresponding to the $^{42}$Ar activity of $1.2^{+0.3}_{-0.5}$ $\mu$Bq per cubic meter of air.

\end{abstract}

\begin{keyword}

\texttt{$^{42}$Ar, low background experiments}

\MSC[2010] 00-01\sep  99-00

\end{keyword}

\end{frontmatter}



\section{Introduction}

\begin{figure}
  \begin{center} 
    \includegraphics[width=10cm]{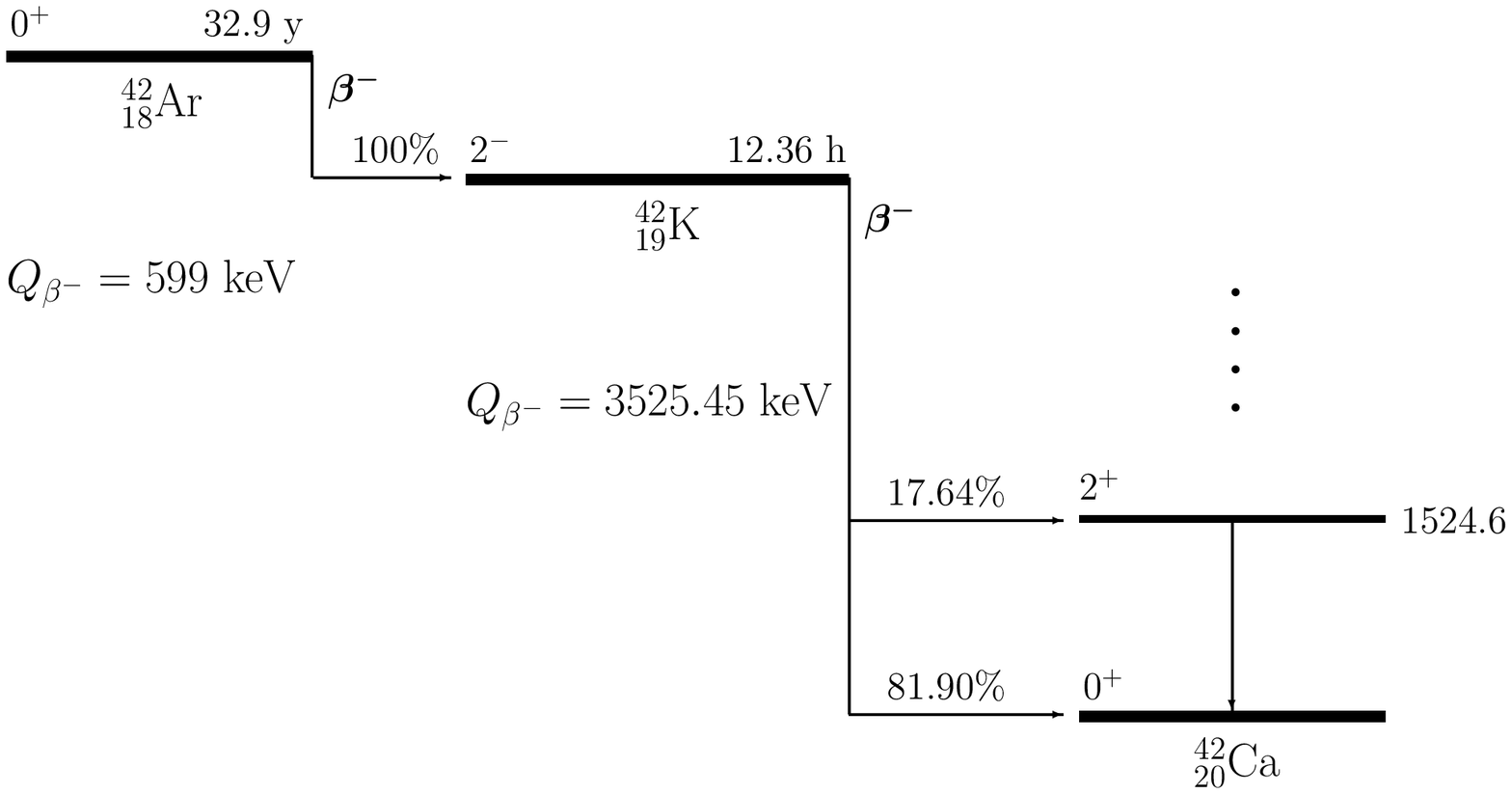} 
  \end{center} 
  \caption{A simplified decay scheme of $^{42}$Ar.}
  \label{fig:ar42_decay_scheme} 
\end{figure}

The long-lived isotope of argon, $^{42}$Ar, is a potential background source in argon-based 
low background detectors. The isotope undergoes a beta decay with a half-life of 32.9 years and the beta decay of 
its daughter isotope,  $^{42}$K, has the maximum electron energy of 3.52 MeV. The decay scheme 
is shown in Figure~\ref{fig:ar42_decay_scheme}.  The potential background problem from $^{42}$Ar 
was first identified  by the ICARUS experiment \cite{BAH86,CEN92}. 
Ray Davis pointed out that a significant amount of $^{42}$Ar could be formed in the Earth's atmosphere 
due to nuclear weapon tests in the upper atmosphere carried out  at the end of the 1950s$-$beginning 
of the 1960s \cite{DAV79}. 

Estimates of the $^{42}$Ar 
concentration in atmospheric argon were carried out in 1995 using the available information on nuclear tests 
\cite{BAR95,CEN95}. During such tests $^{42}$Ar can be produced via a two-step 
neutron capture reaction:

                                $^{40}$Ar(n,$\gamma$)$^{41}$Ar                                                            (1)

                                $^{41}$Ar(n,$\gamma$)$^{42}$Ar                                                            (2)

The above mechanism of $^{42}$Ar production requires a very high neutron flux, the condition fulfilled during nuclear tests in the upper atmosphere. In addition, the half-life of $^{41}$Ar is sufficiently long  ($T_{1/2}  = 1.83$ h) in comparison with a typical time of a nuclear explosion (of the order of milliseconds) to allow an efficient production of $^{42}$Ar. It was shown that the maximum concentration of $^{42}$Ar caused by nuclear tests 
is of the order of $\sim 10^{-22}-10^{-23}$ atoms $^{42}$Ar/atom $^{40}$Ar. 
An alternative mechanism of $^{42}$Ar production was suggested 
in 1997 \cite{PEU97} leading to a substantially higher level of the isotope content in the atmosphere.
In this mechanism $^{42}$Ar is produced by cosmic ray interactions in the upper atmosphere
via the reaction $^{40}$Ar ($\alpha$,2p) $^{42}$Ar.
The corresponding $^{42}$Ar/$^{40}$Ar ratio was estimated to be $\sim 10^{-20}$ atoms $^{42}$Ar/atom
$^{40}$Ar.

The first experimental limit on the content of $^{42}$Ar in atmospheric argon was obtained 
in 1992, $< 10^{-18}$ atoms $^{42}$Ar/atom $^{40}$Ar \cite{ARP92}. 
Later on, a more sensitive estimate came from 
the DBA experiment that ran in the Gran Sasso Underground Laboratory
during 1995-2000 \cite{ASH01,ASH03}. The primary goal of the experiment was 
the search for the double beta decay of $^{100}$Mo using a liquid argon ionization chamber.
The first limit on the $^{42}$Ar content was obtained from a partial data set, $< 6\cdot10^{-21}$
atoms $^{42}$Ar/atom $^{40}$Ar \cite{BAR98}, followed by an updated
limit from the full data set, $< 4.3\cdot10^{-21}$ atoms $^{42}$Ar/atom $^{40}$Ar at 90\% C.L. \cite{ASH03}.
It should be noted that a surplus of events above the expected
background was observed in the DBA experiment in a single electron
energy spectrum. 
However, due to multiple other sources that can contribute to this event topology it was difficult to attribute these events to
$^{42}$Ar. A conservative approach was therefore adopted and
only an upper limit established. 

More recently, the GERDA-I experiment 
carried out an independent measurement of the $^{42}$Ar concentration \cite{AGO14}. 
The aim of the experiment is to search for the neutrinoless double beta decay (0$\nu\beta\beta$)  
of $^{76}$Ge. The experimental signature of  0$\nu\beta\beta$ is a peak at the $Q$-value of the decay (2039 keV). 
The isotope of $^{42}$K is therefore a potential background source for this experiment.

GERDA-I started data taking in 2011. 
High-purity germanium (HPGe) detectors were immersed directly in a vessel with $\sim$ 90 tons of liquid argon. 
The 1525 keV gamma line from $^{42}$K was used to estimate the $^{42}$Ar content. 
The interpretation of the observed count rate in terms of the $^{42}$Ar concentration depends crucially on the assumption 
of the spatial distribution of $^{42}$K ions. It was shown previously (see for example \cite{BAR79}) 
that positive $^{42}$K ions are preferentially formed as a result of the $^{42}$Ar $\beta^{-}$ decay. 
Their spatial distribution in liquid argon is determined by the electric field configuration in the detector.
A high voltage of $\sim$ 4 kV is typically applied to HPGe detectors. As a result, $^{42}$K ions drift in the electric field 
and are accumulated in the vicinity of the detectors leading to a higher counting rate around the 1525 keV gamma line 
than that expected from a uniform distribution of $^{42}$K in liquid argon.

The GERDA collaboration carried out a thorough study of the influence of the electric field on the level of background and methods of its reduction. 
Thin-walled copper containers, mini-shrouds, were constructed around HPGe detectors to isolate them from the main liquid argon volume. 
It resulted in a significant reduction of the background in the 0$\nu\beta\beta$ region and allowed 
the $^{42}$Ar concentration in liquid argon to be estimated, giving a range of  
$(7-12)\cdot10^{-21}$ atoms $^{42}$Ar/atom $^{40}$Ar \cite{AGO14} 
\footnote{The GERDA low-background test facility LArGe was also used to study the $^{42}$Ar background. 
Its concentration was found to be $(2.2 \pm 1.0)\cdot10^{-21}$ atoms $^{42}$Ar/atom $^{40}$Ar. 
The assumption of a uniform distribution of $^{42}$K in liquid argon was used \cite{HEI11}. 
The effect of the electric field from PMTs immersed in the liquid argon volume was not taken into account in the estimation. 
Consequently, the $^{42}$Ar concentration could have been underestimated, which was mentioned in \cite{HEI11}). }. 
Although this value is higher than the limit obtained in \cite{ASH03}, it is worth noting that, within errors, the observed 
effect corresponds to the excess seen in the DBA detector. 
The new analysis of the DBA single electron events presented here has been triggered by the GERDA-I result.

\section{Updated DBA result on $^{42}$Ar concentration}

The data obtained in the DBA experiment \cite{ASH03} have been reanalyzed to estimate the $^{42}$Ar content in the Earth's atmosphere.
The detector was located in the Gran Sasso Underground Laboratory  at a depth of 3500 m of water equivalent. 
The experimental setup consisted of a liquid Ar ionization chamber (Fig. 2) placed in a 15 cm lead passive shielding, 
an Ar purification and delivery system, readout electronics and a data acquisition system. 
The fiducial part of the detector is composed of alternating circular planes of anodes and cathodes 
with Frisch screening grids placed between them. The cathodes are made of a molybdenum foil approximately 50 mg/cm$^2$ thick. 
The chamber contains 14 cathodes, 15 anodes and 28 screening grids. The grid-anode distance is 5.5 mm and the grid-cathode is 14.5 mm. 
The fiducial volume diameter is 30 cm and its height is 56 cm. Each anode is connected to a charge-sensitive preamplifier, 
followed by an amplifier and a shaper. Shaped and unshaped signals are digitized by two 8-bit flash ADC with a 50 ns sampling time. 
Unshaped pulse analysis provides spatial information of the events. 
The trigger for data collection requires that at least one anode signal exceeds the threshold ($\sim$ 600 keV). 
The trigger causes digitized signals from all anodes to be recorded. A more detailed description of the experiment
can be found in \cite{ASH03}.

\begin{figure}

  \begin{center} 

    \includegraphics[width=10cm]{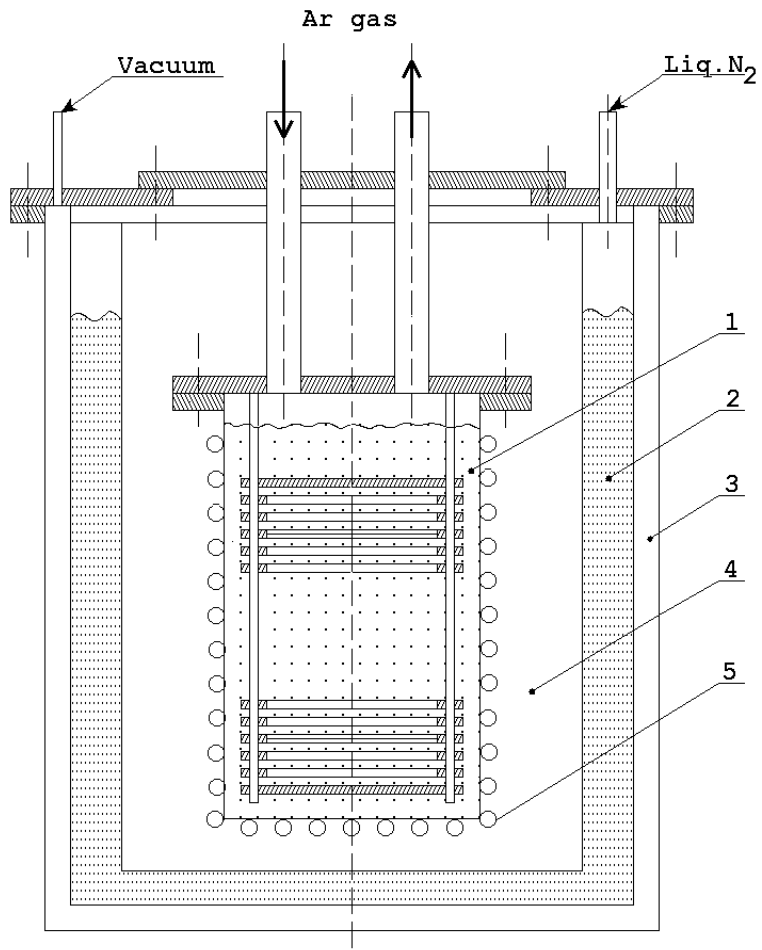} 

  \end{center} 

  \caption{Diagram of the liquid-argon ionization detector used in the DBA experiment: ( 1 ) chamber filled with liquid argon; ( 2 ) thermostat volume filled with liquid nitrogen; ( 3 ) vacuum volume of the thermostat; ( 4 ) space filled with nitrogen vapor; and ( 5 ) system of heaters.}

  \label{fig:det} 

\end{figure}

During the data taking period the electric field was 1.9 kV/cm in the cathode-grid gap and 4 kV/cm in the anode-grid gap. 
The detector energy scale was calibrated using $^{22}$Na (E$_{\gamma}$ = 
1275 keV) and $^{88}$Y ((E$_{\gamma}$ = 1836 keV) radioactive
sources (see description in \cite{ASH03}). The energy resolution of the chamber was $\sim 6$\%
(FWHM) at the $0\nu\beta\beta$ transition energy of $^{100}$Mo, 3 MeV. 

A single electron signature is used to select candidate events from $^{42}$K decays.
This signature requires a single energy deposition detected on one of the anodes
with no signals on all other anodes.
The 3.0-3.5 MeV energy interval of the single electron spectrum accumulated over 2706 hours 
is used to determine the $^{42}$Ar concentration in liquid argon. 

\begin{figure}
  \begin{center} 
    \includegraphics[width=10cm]{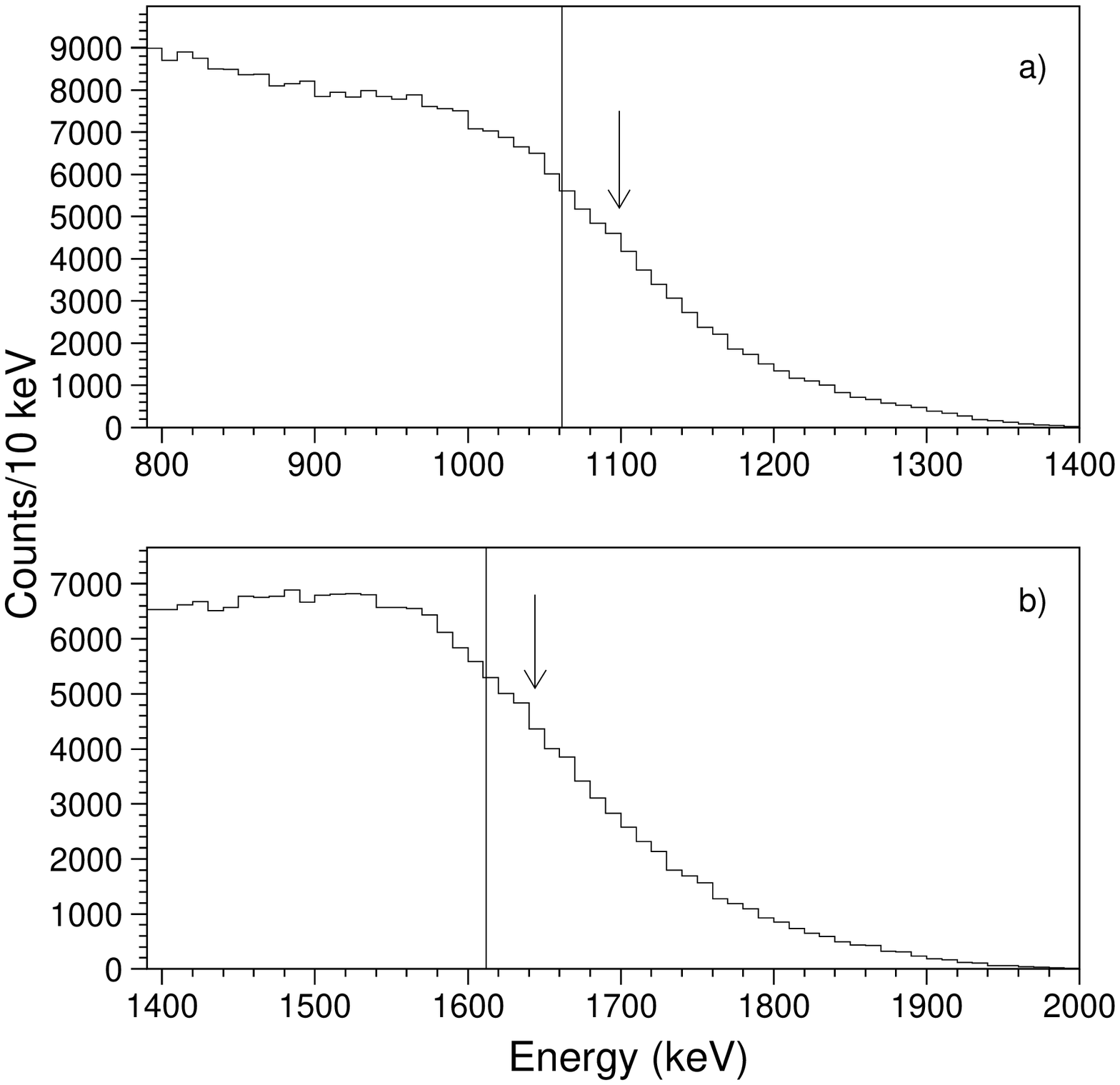} 
  \end{center} 
  \caption{Simulated energy spectra of single electrons in the DBA liquid argon detector 
  from the sources of $^{22}$Na (a) 
and $^{88}$Y (b). Arrows show inflection points at 1099 keV (a) and at 1644 keV (b). 
Vertical lines correspond to the maximum energy of Compton electrons, 1061.70 keV and 1611.78 keV, respectively.}
  \label{fig:compton} 
\end{figure}

The new analysis of the experimental data has shown a small shift in the energy scale of the calibration spectra 
that was unaccounted for in the original analysis. A 5th-degree polynomial is used to fit the high energy edge of the Compton spectrum. 
The inflection point of the fitting function corresponds to the maximum energy of Compton electrons. 
However, multiple interactions of $\gamma$ rays in a single cathode-grid volume shift the inflection point to higher
energies. This effect is illustrated in Figure 3. 

Decays of $^{208}$Tl from materials surrounding the detector are used as an additional
point of the energy calibration
that takes into account multiple gamma-ray interactions in a single cathode-grid gap. 
Figure 4 shows the fit of simulated $^{208}$Tl events in materials surrounding
the fiducial volume of the detector to the experimental single energy spectrum around 2381.76 keV. 
This energy corresponds to the maximum energy of Compton electrons from the 2614 keV line of $^{208}$Tl. 

The energy scale correction has been applied to the single electron energy spectrum, which is shown in Figure~\ref{fig:singleE_spectrum}.

\begin{figure}
  \begin{center} 
    \includegraphics[width=10cm]{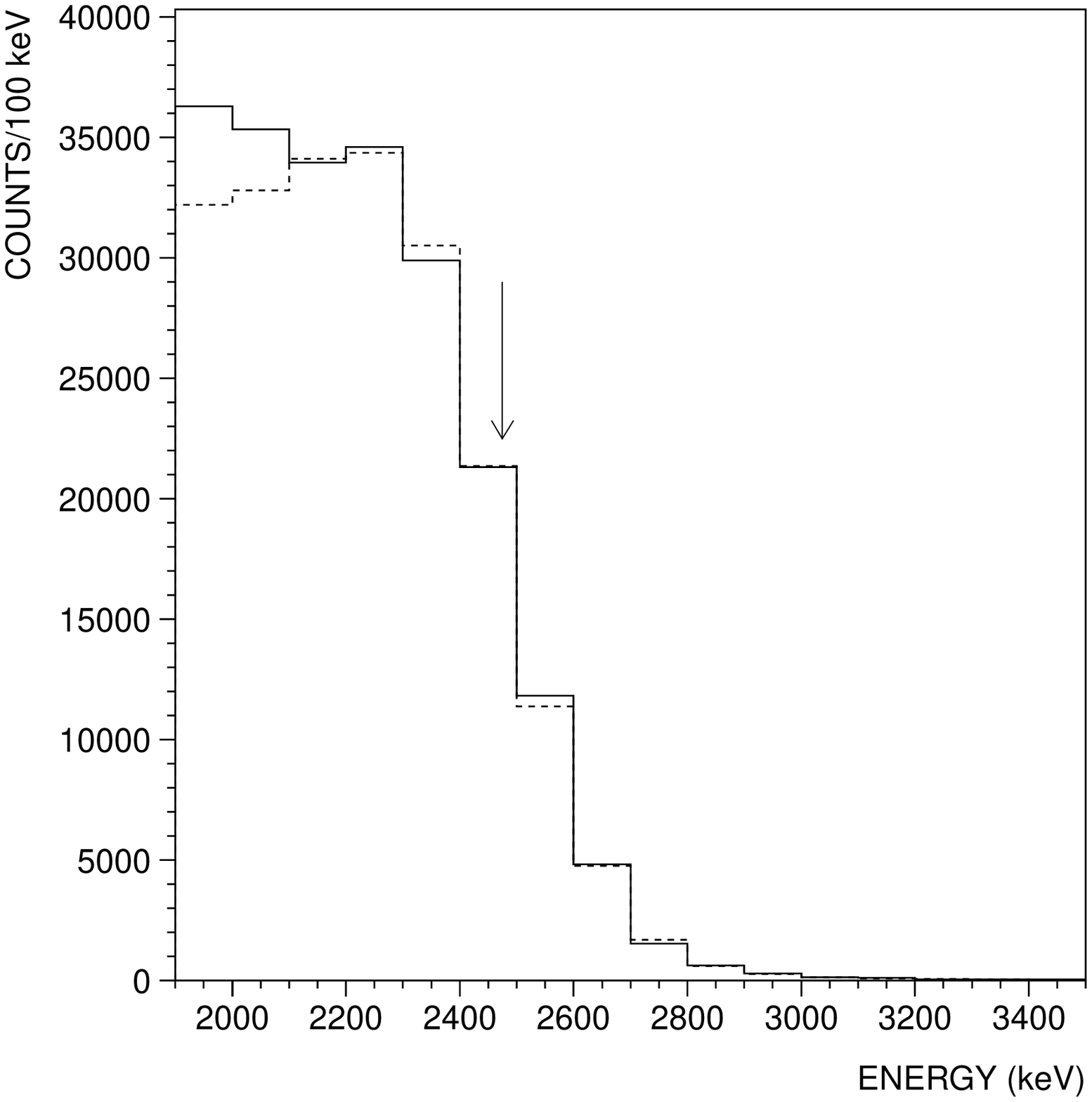} 
  \end{center} 
  \caption{High energy part of the single electron spectrum observed in data (solid line). The dashed line shows the simulated $^{208}$Tl events in detector materials fitted to the experimental data.
The inflection point of the experimental spectrum is shown by the arrow at 2475 keV.}
  \label{fig:tl_titan} 
\end{figure}

\begin{figure}
  \begin{center} 
    \includegraphics[width=10cm]{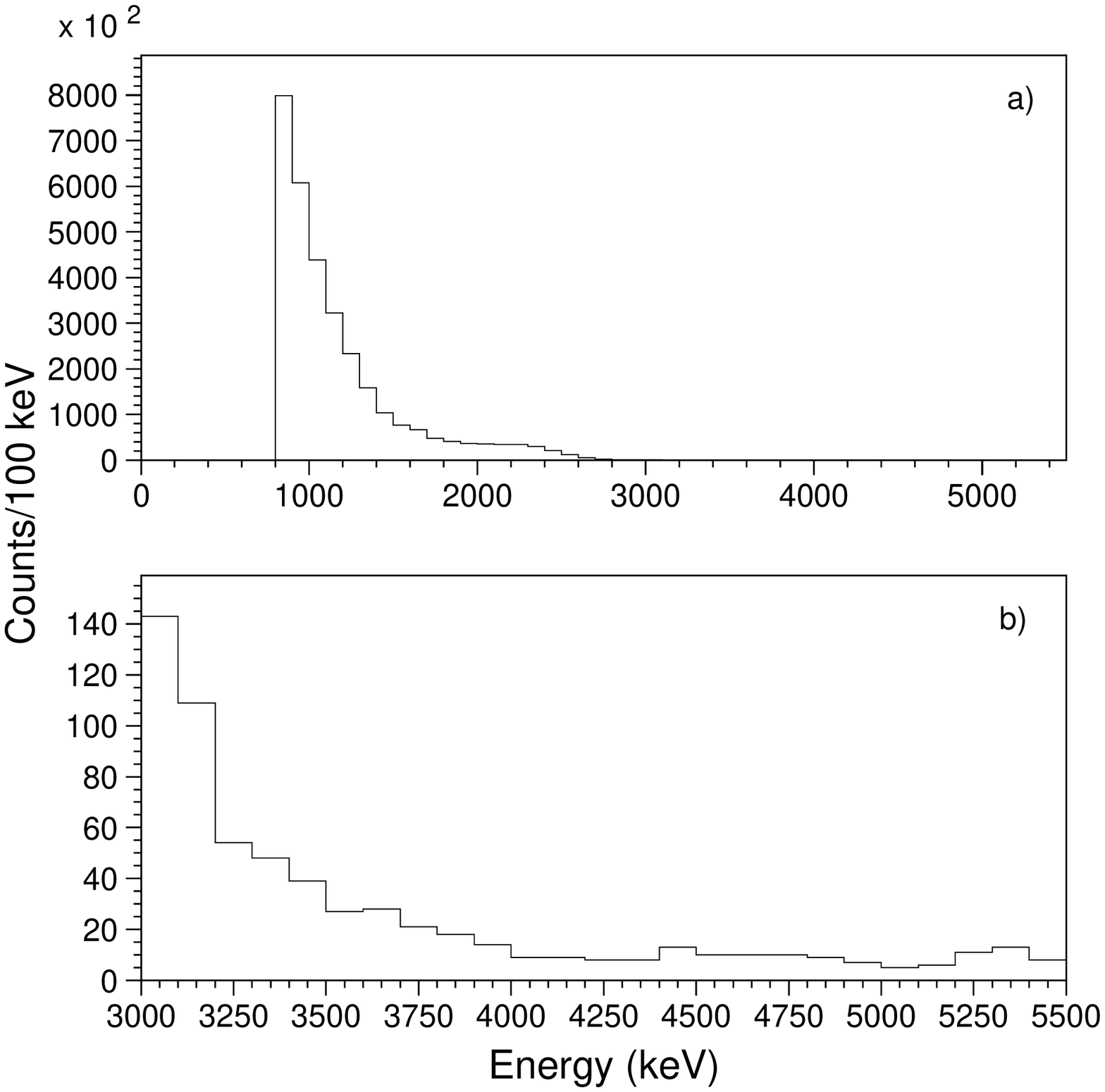} 
  \end{center} 
  \caption{Single electron energy spectrum in the region of 0-5.5 MeV (a) and 3.0-5.5 MeV (b).
 The measurement time is 2706 hours.}
  \label{fig:singleE_spectrum} 
\end{figure}

MC simulations have been carried out to evaluate the detection efficiency in the 3.0-3.5 MeV energy interval for two possible distributions of  $^{42}$K in liquid argon.
In the first case, positive ions of $^{42}$K created in the $\beta$ decay of $^{42}$Ar are assumed 
to maintain their charge for a sufficiently long time to be transported to negatively charged cathodes of the detector and 
undergo their $\beta$ decay from there, as was shown in \cite{BAR79}. In the second case, a uniform distribution 
of $^{42}$K in the liquid argon volume of the detector is assumed. The efficiency is calculated to be 0.87\% and 1.07\% 
in the first and second case respectively.  The calculations below assume the accumulation of positive $^{42}$K ions on negatively 
charged cathodes of the chamber. However, a uniform distribution of neutral $^{42}$K atoms in the detector volume 
will not alter the final result by more than 15\%.  Recently, EXO-200 reported the fraction of 
 $^{214}$Bi$^+$ ions in the $^{214}$Pb $\beta$ decay (in liquid Xe) to be $(76.4 \pm 5.7)$\%, 
 while the remainder of $^{214}$Bi atoms are neutral \cite{ALB15}. Assuming the same ratio of 
 positive to neutral $^{42}$K atoms from the $^{42}$Ar decay (in liquid Ar) the above efficiency
  is estimated to be 0.92\%. Ultimately, the uncertainty on the efficiency due to the distribution of  the  $^{42}$K ions
  in the volume of the liquid argon detector is included in the systematic uncertainty on the $^{42}$Ar concentration calculated below. 

Table 1 shows possible other contributions to the single electron events in the 3.0-3.5 MeV energy interval 
which are treated as background for the $^{42}$Ar measurement. The contribution from $n\gamma$ reactions are evaluated 
by extrapolating the flat part of the spectrum at high energies down to lower energies.  The external background from $^{214}$Bi is 
estimated using the available information about the radiopurity of titanium ($< 30$ mBq/kg), 
which was the primary material of the detector vessel. 
The $^{208}$Tl external background is estimated by fitting the Compton edge of the single electron spectrum around 2400 keV (see Figure~\ref{fig:tl_titan}),
and evaluating its contribution to the single electron events in the 3.0-3.5 MeV energy region. 

\begin{table}[ht]

\label{Table1}

\caption{Possible background contributions (in addition to $^{42}$Ar) to single electron events in the energy region of 3.0-3.5 MeV.
The limits are given at 95\% C.L.}

\vspace{0.5cm}


\begin{center}

\begin{tabular}{cc}

\hline

Background & Number of events \\

\hline

$\gamma$-rays from neutrons & 66 \\

$^{208}$Tl (external)  & 87 \\ 

$^{214}$Bi (external) & $<$ 0.4 \\ 

$^{208}$Tl (Mo cathodes) &  $< 89$ \\

$^{214}$Bi (Mo cathodes) & $< 10$ \\

\hline

\end{tabular}

\end{center}

\end{table}

Limits obtained from screening measurements with low-background HPGe detectors are used to constrain the background 
from the internal contamination inside the molybdenum cathode foils (the specific activities of 
$^{208}$Tl and $^{214}$Bi are $<$ 1.6 mBq/kg and $<$ 10 mBq/kg, respectively). 
Taking into account the above background sources the contribution of $^{42}$Ar to the single electron events 
observed in the 3.0-3.5 MeV interval over the period of data taking is estimated to be $240 \pm 23(stat) ^{+55}_{-110}(syst)$ events.  
The main source of the systematic error is a relatively weak constrain on $^{208}$Tl in the Mo cathode foils
(see Table 1). Other sources of the systematic error include the energy scale calibration, the $^{42}$K ion distribution in
liquid argon and its effect on the detection efficiency. 

The argon used in the detector had been produced $\sim$ 10 years before the measurements were carried out. 
Thus, given the half-life of $^{42}$Ar (32.9 years) the concentration of the isotope in the detector's volume is 20\% lower than that in the Earth's atmosphere. 

Using the above parameters and accounting for the total fiducial Ar mass of 52 kg, the concentration of  $^{42}$Ar in the Earth's atmosphere can be estimated as 
$6.8^{+1.7}_{-3.2}\cdot10^{-21}$ atoms of $^{42}$Ar/atom of $^{40}$Ar, which corresponds to the  $^{42}$Ar activity of 
$1.2^{+0.3}_{-0.5}$ $\mu$Bq/m$^3$ of air. This leads to a specific activity of freshly produced liquid argon of $68^{+17}_{-32}$ $\mu$Bq/kg, in agreement with the earlier GERDA-I result \cite{AGO14}. We also note that the presented result is consistent with the mechanism of $^{42}$Ar production suggested in \cite{PEU97}.

\section*{Acknowlegements}

This work was supported by RFBR under the research project No. 15-02-02919.

\end{document}